\documentclass[useAMS,usenatbib]{mn2e}
\usepackage{graphicx}
\usepackage{amsmath}
\usepackage{makecell}
\usepackage{subcaption}
\usepackage{gensymb}
\usepackage[normalem]{ulem}
\usepackage{comment}
\usepackage{macros}
\usepackage[usenames,dvipsnames,svgnames,table]{xcolor}
\usepackage{hyperref}
\definecolor{darkblue}{rgb}{0.0,0.0,0.3}
\hypersetup{colorlinks,%
            linkcolor=darkblue,urlcolor=darkblue,
            anchorcolor=darkblue,citecolor=darkblue}
\usepackage{txfonts}
\DeclareSymbolFont{cmletters}{OML}{cmm}{m}{it}
\DeclareMathSymbol{v}{\mathalpha}{cmletters}{"76}
\newcommand{\RedeclareMathOperator}[2]{\renewcommand{#1}{}\let#1\relax\DeclareMathOperator{#1}{#2}}

\newcommand\simless\lesssim
\newcommand\simgreat\gtrsim

\newcommand\highlight[1]{#1}

\title[Precessing Tilted Disc-Jet Simulations]{Formation of Precessing
  Jets by Tilted Black Hole Discs in 3D General Relativistic MHD Simulations}
\author[Liska, Hesp, Tchekhovskoy, Ingram, van der Klis \& Markoff]{M. Liska$^1$, C. Hesp$^{1,2}$, A. Tchekhovskoy$^{3,4,5,6}$, A. Ingram$^1$, M. van der Klis$^1$, S. Markoff$^{1,2}$\\
$^{1}$Anton Pannekoek Institute for Astronomy, University of Amsterdam, Science Park 904, 1098 XH Amsterdam, The Netherlands\\
$^{2}$Gravitation Astroparticle Physics Amsterdam (GRAPPA) Institute, University of Amsterdam, Science Park 904, 1098 XH Amsterdam, The Netherlands\\
$^{3}$Center for Interdisciplinary Exploration \& Research in Astrophysics (CIERA),
Physics \& Astronomy, Northwestern University, Evanston, IL 60202, USA\\
$^{4}$Departments of Astronomy and Physics, Theoretical Astrophysics Center, University of California Berkeley, Berkeley, CA 94720-3411, USA \\
$^{5}$Lawrence Berkeley National Laboratory, 1 Cyclotron Rd, Berkeley,
CA 94720, USA \\
$^{6}$Kavli Institute for Theoretical Physics, Kohn Hall, University
of California at Santa Barbara, Santa Barbara, CA 93106
}

\begin{document}

\date{Accepted. Received; in original form}
\pagerange{\pageref{firstpage}--\pageref{lastpage}} \pubyear{2017}

\maketitle
\label{firstpage}

\begin{abstract}
Gas falling into a black hole (BH) from large distances is unaware
of BH spin direction, and misalignment between the
accretion disc and BH spin is expected to be common. However, the
physics of tilted discs (e.g., angular momentum transport and
jet formation) is poorly understood.  Using our
new GPU-accelerated code H-AMR, we performed 3D
general relativistic magnetohydrodynamic simulations of tilted thick
accretion discs around rapidly spinning
BHs, \highlight{at the highest resolution to date. We explored the limit where
disc thermal pressure dominates magnetic
pressure, and showed for the first time that, for different} magnetic field
strengths on the BH, these flows launch magnetized relativistic jets propagating along the rotation axis of
the tilted disc (rather than of the BH). If
strong large-scale magnetic flux reaches the BH, it bends the inner few gravitational radii of
the disc and jets into partial alignment with the BH spin.
On longer time scales, the simulated disc-jet system as a whole undergoes
Lense-Thirring precession and approaches alignment, 
demonstrating for the first time that jets can be used as probes of disc
precession. When the disc turbulence is well-resolved, our isolated
discs spread out, causing both the alignment and precession to slow
down.

\end{abstract}


\section{Introduction}
\label{sec:introduction}

The angular momentum of matter accreting onto a spinning black hole
(BH) is expected to often be misaligned with the BH spin, in wide
range of systems including X-ray
binaries, active galactic nuclei (AGN), tidal disruption events (TDEs), \highlight{and BH-neutron star mergers},
with observations indicating the presence
of such discs in some systems \citep[e.g.,][]{hjellming95,caproni06}. 

Disc tilt has been invoked to explain quasi-periodic oscillations
(QPOs) in BH systems \citep{stella98,ingram09} and variability in the
jet orientation
\citep{2012PhRvL.108f1302S,2012ApJ...747...63A,2013AJ....146..120L,2016AJ....152...12L,tmgk13}. 
Without an accurate theoretical
description of tilted accretion flows, 
3D MHD simulations in general
relativity (GR) are an excellent tool for understanding these
important systems.

An effect of crucial importance for tilted accretion is \emph{Lense-Thirring precession} \citep[LT;][]{lense18}. In GR, massive rotating objects distort nearby inertial frames (``frame-dragging"). Such twisting induces nodal precession in test particles on \emph{tilted orbits}, depending on their distance to the central object (precession frequency $\propto 1/r^3$).
We characterise warps in the disc in terms of the precession angle $\mathcal{P}(r)$ with respect to the initial disc orientation and the tilt angle $\mathcal{T}(r)$ with respect to the BH spin. MHD effects allow perturbations in $\mathcal{T}(r)$ and $\mathcal{P}(r)$ to travel radially, affecting the overall disc behaviour. 

Disc warps can travel by means of pressure waves and viscosity
(parametrized by the viscosity parameter $\alpha$), which is generated by
magnetic turbulence. In thin high-viscosity discs, with scale height
$H/R<\alpha$, pressure waves are damped and warps travel by viscous
diffusion
\citep{papaloizou83}. %
Here, we work in the opposite limit of
thick low-viscosity discs, with $H/R>\alpha$. In such discs, pressure
waves can travel freely at about half the sound speed
\citep{papaloizou95}. In this wave-like limit, $\mathcal{T}(r)$
oscillates in radius at $r\lesssim 20 \: r_g$
\highlight{\citep{demianski97,ivanov97,lubow00,lubow02}}. This prediction was confirmed in
\highlight{simulations of tilted accretion discs using GR hydrodynamics \citep[][]{mewes16} and GRMHD \citep{fragile05,fragile07}. The latter}
simulations also showed LT precession of the disc around the BH at a
constant rate (constant ${\rm d}\mathcal{P}/{\rm d}t$), with the disc behaving
as a rigid body. 
Interestingly, the precession frequency
was found to be consistent with observed Type-C QPOs.

On the other hand, jets can strongly affect the accretion flow and its orientation: when the magnetic flux on the BH is strong enough to obstruct
the inner disc infall \citep{nia03}, the associated jets extract large amounts of
rotational energy from the BH and the disc
\citep{tch12proc,2015ASSL..414...45T} and compress the disc vertically
\citep{tch11,tchekhovskoy12,mckinney12}. In this so-called
\emph{magnetically arrested disc} (MAD) regime, the jets can
force the inner parts of tilted thick,
radially extended accretion discs to align with
the BH spin \citep{mtb13,polko17}.  However, the behaviour of jets
produced by tilted discs with smaller magnetic
fluxes, in the so-called \emph{standard and normal evolution} (SANE, \citealt{2012MNRAS.426.3241N}) regime, and/or smaller
radial extents remains completely unexplored.

Here we study tilted thick disc-jet systems for a range of magnetic
field strength and with disc size changing substantially over time, %
 using first-principles GRMHD
simulations. %
We describe our numerical method in Sec.~\ref{sec:h-amr-code} and our numerical setup in Sec.~\ref{sec:numerical-models}. We present our results in Sec.~\ref{sec:results} and conclude in Sec.~\ref{sec:Discussion}.

\section{H-AMR (``hammer'') code}
\label{sec:h-amr-code}

\begin{table}
\centering
\label{Parameters}
\begin{tabular}{@{}l@{\;\;}l@{\;\;}c@{\;\;}c@{\;\;}c@{\;\;}c@{}}
\hline
Model & \hfill $a$\hfill\hbox{} & $r_{\rm in}\ [r_g]$ &$r_{\rm max}\ [r_g]$ & $\mathcal{T}_{\rm init}\ [{\rm deg}]$\\
\hline
All & \hfill 0.9375 \hfill\hbox{} &12.5 & 25&30\\
\\
\hline
Short &Full model &B-flux   & Resolution,  & Q-factor & $t_{\rm sim}$\\
name & name  & strength & $N_{r}\times N_\theta\times N_\varphi$
                                                        & $Q_{r},
                                                          Q_\theta,
                                                          Q_\varphi$ &$[10^5t_g]$ \\
\hline
\href{https://www.youtube.com/watch?v=UDuD9IZmcvg&index=2&list=PL39mDr1uU6a5RYZdXLAjKE1C_GAJkQJNv}{S-R}&S25A93&Strong&$448\times144\times240$&($9,9,32$)   & $1.2$\\
\href{https://www.youtube.com/watch?v=Jiw1qj6veL0&index=4&list=PL39mDr1uU6a5RYZdXLAjKE1C_GAJkQJNv}{S-HR}&S25A93HR&Strong&$896\times288\times480$&($30,28,80$) & $0.75$\\
\href{https://www.youtube.com/watch?v=lWjowRhyNk8&list=PL39mDr1uU6a5RYZdXLAjKE1C_GAJkQJNv&index=6}{W-U}&S25A93WLR&Weak&$448\times144\times240$&($2.5,3,16$)   & $1.2$\\
\href{https://www.youtube.com/watch?v=O2-LcxY4E5w&index=3&list=PL39mDr1uU6a5RYZdXLAjKE1C_GAJkQJNv}{W-R}&S25A93W&Weak&$896\times288\times480$&($20,18,53$)   & $1.2$\\
\href{https://www.youtube.com/watch?v=Aq_McDaByGk&index=5&list=PL39mDr1uU6a5RYZdXLAjKE1C_GAJkQJNv}{W-HR}&S25A93WHR&Weak&$1792\times576\times960$&($59,55,138$) & $0.3$\\
\hline
\end{tabular}
\caption{ %
  [top panel:] Parameters common to all models: BH spin $a$,
  the radii of the torus inner edge $r_{\rm in}$ and pressure maximum
  $r_{\rm max}$, the initial tilt angle $\mathcal{T}_{\rm init}$, and
  simulation duration $t_{\rm sim}$.
  [bottom panel:] Short and full model name (i.e., S25A93 stands for
  size and BH spin, W for weakly magnetised, LR or HR for low or high
  resolution), the strength of the magnetic flux, the resolution
  $N_{r,\theta,\phi}$, and the quality factor $Q_{r,\theta,\phi}$ (the number of cells per
  MRI fastest growing wavelength in $r-$, $\theta-$, and $\phi-$directions) at $t =
  5\times 10^4 \: t_g$. \highlight{Short names of the models link to 3D animations in a \href{https://www.youtube.com/playlist?list=PL39mDr1uU6a5RYZdXLAjKE1C_GAJkQJNv}{YouTube playlist}.}}
\label{tab:models}
\end{table}
 
We use a new  massively parallel 3D GRMHD code H-AMR (pronounced
``hammer'') accelerated by Graphical
Processing Units (GPUs). We developed H-AMR based on a 2D serial open-source code
HARM2D \citep{gammie03,nob06}. H-AMR performs $10$ times faster on a GPU than on a 16-core CPU. H-AMR
is parallelised via MPI with domain decomposition and scales well to
thousands of GPUs, achieving weak scaling efficiency of $85$\% on
$4096$ GPUs for a tile size of $100^3$ cells on the Blue Waters
supercomputer \hbox{\highlight{(see Supporting
  Information [SI] Section 2.1).}}

H-AMR features a staggered grid for constrained transport of magnetic
field \citep{gar05}, adaptive mesh-refinement (AMR, \highlight{not used adaptively here}), and a locally
adaptive time step \highlight{(details
to be described in future work)}. These specialized features 
substantially speed up the simulations, which we carry out on a
spherical polar grid \highlight{(see also SI Section 2.3).} We use outflow boundary conditions in the $r-$,
transmissive polar boundary conditions in the $\theta-$, and periodic
boundary conditions in the $\phi-$direction. The radial grid is
uniform in $\log r$ and extends from just inside of the event horizon
out to $10^5r_g$, where $r_g=GM/c^2$ is the gravitational
radius, such that the outer boundary is
causally disconnected over the simulation duration, $\gtrsim10^5 t_g$, where $t_g = r_g/c$. \highlight{If our grid were uniform in 
$\phi$ and $\theta$}, cells would become prohibitively small near the polar
axis. To mitigate this problem, we adopt the approach of %
\citet{tch11} of stretching the polar cells in $\theta$ and, additionally,
use $2{-}4$ layers of static mesh-refinement to decrease the
$\phi$-resolution, giving a speedup by a factor of $4{-}16$ \highlight{(see SI Sections 2.2--2.3)}.
Local adaptive time-stepping between the tiles gives an additional 
speed-up by a factor of $2{-}3$. \highlight{In practice, 
these advanced features give a speedup by a factor of $10{-}30$
(see SI Section 2.1)}.
The high speed of the code allows us to study tilted discs
at much higher resolution and over longer durations than was possible
until now, as required to handle the large dynamic range necessary to study tilted accretion and jets in 3D.

\section{Numerical models}
\label{sec:numerical-models}
Our simulations start with a hydrostatic torus
\citep{fishbone76} for a BH spin $a=0.9375$ \highlight{and use a
Kerr-Schild foliation}. We use an
ideal gas equation of state, $p_{g}=(\gamma-1)u_{g}$, where $p_{g}$
and $u_{g}$ are the thermal pressure and energy density, and
$\gamma = 5/3$. We place the torus inner edge at
$r_{\rm in} = 12.5r_g$ and density maximum $\rho_{\rm max}=1$ at
$r_{\rm max} = 25r_g$. We tilt the torus relative to the BH spin by
an angle, $\mathcal{T}_{\rm init} = 30\degree$. In contrast to earlier
work in which the BH spin was tilted with respect to the grid
\citep{fragile05,fragile07,fragile09b,mtb13}, we tilt the disc itself
and leave the BH spin pointed along the polar axis, because an axisymmetric metric lowers the memory
footprint.

We carried out five production models listed in
Table~\ref{tab:models}, each differing in the amount of the initial
magnetic flux and resolution. For our strongly magnetized models (denoted with `S-'), we insert in the torus a large
magnetic field loop, described by the magnetic vector potential
$A_{\phi}\propto(\rho-0.05)^{2}r^{3}$, where $\rho$ is the rest mass
density. %
For our weakly magnetized models (denoted with W-), we use a smaller loop, $A_{\phi}\propto(\rho-0.05)$.  We simulate the two models, \hbox{S-R} and \hbox{W-R}, using sufficiently high
resolutions (resolved, denoted with \hbox{`-R'}; see Table~\ref{tab:models}) to resolve the
magnetorotational instability \citep[MRI;][]{balbus91} that fuels
magnetic turbulence. We simulate models \hbox{S-HR} and \hbox{W-HR} at twice the resolution in all three dimensions to check for convergence (highly resolved, denoted with \hbox{`-HR'}). Finally, we set up a fifth model, \hbox{W-U} to be physically
identical to model \hbox{W-R} but with only half the resolution in all three dimensions (under-resolved, denoted with -U; see
Table~\ref{tab:models}), in order to investigate the effect of
under-resolving the MRI. %

We set the numerical resolution, $N_{r}\times N_\theta\times N_\phi$
(see Table~\ref{tab:models}), to attain a near-unity cell aspect ratio everywhere. We
normalise the magnetic pressure, $p_{B}$, such that it is
subdominant compared to $p_{\rm g}$, by setting $p_{\rm g,max}/p_{\rm B,max} = 100$. %
Following the approach of
\citet{2017MNRAS.467.3604R}, we enforce throughout the simulations that 
$\rho c^2\ge \max(p_{\rm B}/50,10^{-6}c^2(r/r_g)^{-2})$ and
$u_{\rm g} \ge \max(p_{\rm B}/150,10^{-7}c^2(r/r_g)^{-2\gamma})$, 
approximating physical processes
that mass-load relativistic jets at their base, where $p_{\rm B} \gg \rho c^2$. 

We note that using a spherical polar grid to simulate tilted disc-jet
systems is challenging, because improper treatment of the transmissive
polar boundary condition can lead to numerical artifacts. To demonstrate that the disc
and jets pass freely through the polar coordinate singularity, we
carried out test simulations using the parameters of model
S-R \highlight{(see SI Section 2.4)}. Firstly, we verified that the two equivalent ways of simulating a
tilted system -- i.e., separately misaligning (i)~the BH ($a =
0.9375$) or (ii)~the disc by $30$ degrees relative to the grid  -- do
indeed give consistent profiles for $\mathcal{T}(r)$ and $\mathcal{P}(r)$. Secondly, we
verified that the simulation outcome of a BH-torus system is
independent of its orientation relative to the grid. Namely, we simulated an aligned BH-torus system oriented
along the polar axis and compared it to that oriented at 30 degrees
relative to the polar axis, for two values of the BH spin ($a = 0$ and
$a = 0.9375$). Please see SI Sec. 2.4 for more detail.

\begin{figure}
\begin{center}
\includegraphics[width=\linewidth,trim=0mm 0mm 14mm 0,clip]{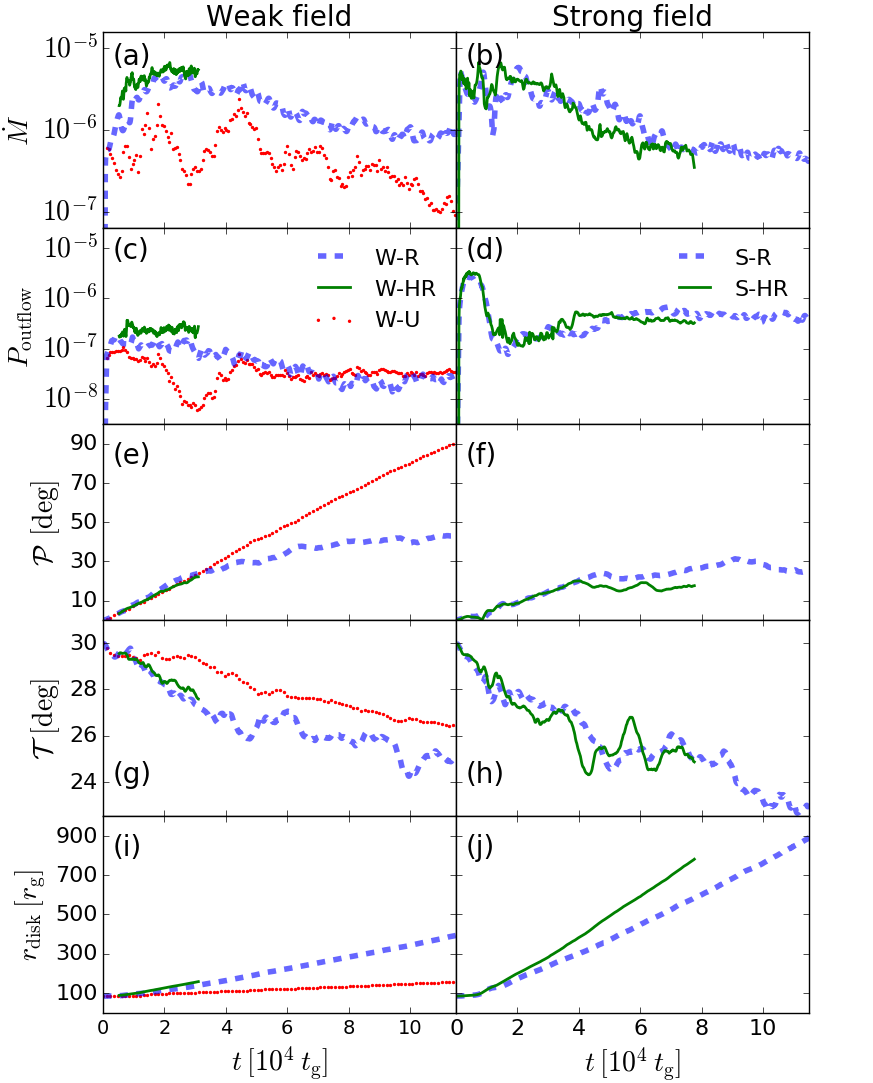}
\end{center}
\caption{%
  Time dependence of mass
  accretion rate $\dot{M}$ in units of $M_{\rm init}/t_g$ [panels
  (a-b)], where $M_{\rm init}$ is the initial torus mass;
  outflow power $P_{\rm outflow}$ in units of
  $M_{\rm init}c^2/t_g$ [panels (c-d)]; precession angle $\mathcal{P}$
  [panels (e-f)] and tilt angle $\mathcal{T}$ [panels (g-h)], where both angles
  measured over the radial interval $50{-}150\: r_g$; and disc radial extent $r_{\rm disc}$
  [panels (i-j)], 
  which is the average radius, in the plane of the disc, weighted by
rest mass density. %
  The left and right columns show respectively the
  weak (W-) and strong (S-) magnetic field models, for the resolved
  runs (-R, thick dashed blue lines), highly resolved runs (-HR, solid
  thin green lines), and under-resolved run (W-U, dotted red
  lines). All variables show approximate convergence between
  the resolved and highly resolved runs. Both differ greatly from the
  under-resolved run (W-U).  }
\label{fig:time}
\end{figure}

\section{Results}
\label{sec:results}
In all five models (Table~\ref{tab:models}), magnetic turbulence
develops and the gas reaches the BH as seen in
Figure~\ref{fig:time}(a-b).  Accretion occurs via two polar plunging
streams, consistent with the findings of \citet{fragile07} who studied
the dynamics of tilted accretion flows with weak magnetic flux at a
resolution similar to our model W-U. As Figure~\ref{fig:time}(c-d)
shows, each of our models produces an
energetically-significant outflow of energy, with power $P_{\rm
  outflow}\equiv \dot{M}c^2-\dot{E} \gtrsim (0.1{-}1)\dot{M}c^2$. Here
$\dot{M}$ and $\dot{E}$ are mass and total energy accretion rates onto the BH. \highlight{For all five models, 3D animations are available (see SI or \href{https://www.youtube.com/playlist?list=PL39mDr1uU6a5RYZdXLAjKE1C_GAJkQJNv}{this YouTube playlist}).} 

\begin{figure}
\includegraphics[width=1.0\linewidth]{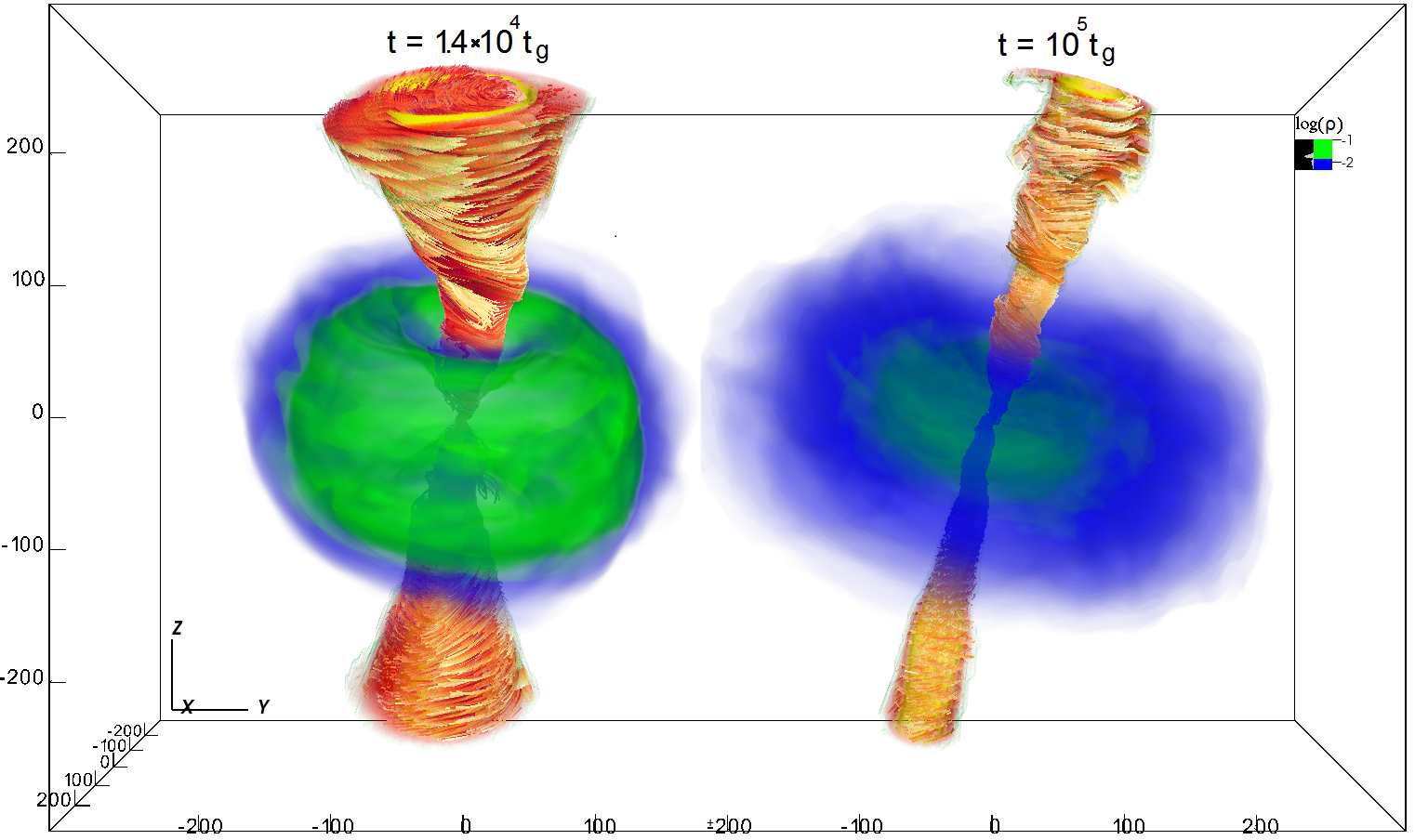} 
\caption{ Volume rendering of density [$\log(\rho)$ in blue and green,
  indicating the disc] and 
  jet magnetic field lines coloured by the scaled magnetic energy density [$p_B r / 
\rho c^2
  > 0.5$ in red and yellow, indicating the jets] for model
  \hbox{W-R} at $1.4\times 10^4 \: t_g$ (left) and $10^5 \: t_g$ (right). The
  magnetic field lines in the jets are shown with yellow-red lines,
  threading the entire volume rendering of the jet. We measure
  distances in units of $r_g$. The disc-jet system precesses as a
  whole around the BH spin vector, which is vertical in the figure. As
  the simulation progresses, the disc spreads outward, its density
  profile flattens, and the distance between the high and low density
  regions increases. %
}
\label{fig:visual}
\end{figure}
Figure~\ref{fig:visual} shows a volume rendering for model W-R after $1.4 \times 10^4 \: t_g$ and $10^5 \: t_g$. 
Model S-R looks similar, but its jet has a larger opening angle
reflecting the stronger outflow power than in model W-R [see Figure~\ref{fig:time}(b)].
In all our models, we find that the jets
are connected by magnetic field lines to the event horizon (see
Figure~\ref{fig:visual}) and
therefore appear to be powered by the extraction of BH rotational
energy via the Blandford-Znajek \citep[BZ;][]{blandford77} mechanism.
They reach relativistic Lorentz factors $\gamma\gtrsim10$ at
$r\lesssim 200 r_g$, similar to 2D jets
\citep{bes06,mckinney06,tch08}.

Comparison of the two panels in Figure~\ref{fig:visual} shows that the
disc orientation changes over time, i.e., the tilted disc
\emph{precesses}. This is also seen in  Figure~\ref{fig:time}(e-f)
through the increase in $\mathcal{P}$. We measure $\mathcal{P}$ and $\mathcal{T}$ of the disc by calculating its
  angular momentum vector \highlight{\citep{nelson00, fragile05}} as a
  function $r$. [We measure the location of 
  of the jet by isolating the highly magnetized region (using $p_B
  > \rho c^2/ 2r$) and weighing its position by the magnetic pressure
  on spherical shells (see SI Section 3).] In addition to precession, the
disc also aligns with the BH spin, as seen in Figure~\ref{fig:time}(g-h)
through the decrease in $\mathcal{T}$. In the simplest case, model \hbox{W-U}, the disc
precesses at a constant rate, ${\rm d}\mathcal{P}/{\rm d}t={\rm
  constant}$, as expected for a misaligned disc angular
momentum vector under the action of a constant LT
torque \citep[see][]{fragile07}. \highlight{LT precession also induces
  a twist in the innermost disc. That effect remains relatively
  constant over time (in agreement with \citealt{fragile07}), allowing for solid--body-like precession of the
  whole system.}

Figure~\ref{fig:time}(e-h) shows that in our models with well-resolved MRI,
\hbox{S-R} and \hbox{W-R}, the alignment and
precession slow down over time. 
Indeed, well-resolved MRI is crucial for capturing disc angular momentum
transport and outward expansion [see Figure~\ref{fig:time}(i-j)] that
brings the disc angular momentum out of reach of jet and LT
torques (both are stronger near the BH) and impedes the alignment
and precession. 
The
approximate agreement between the \hbox{-R} and \hbox{-HR} models
(with numerical resolutions in every dimension
different by a factor of $2$)
indicates that our results are reasonably converged with the numerical
resolution.

\begin{figure}
\centering\includegraphics[width=\linewidth,trim=8 0 40 70,clip=true ]{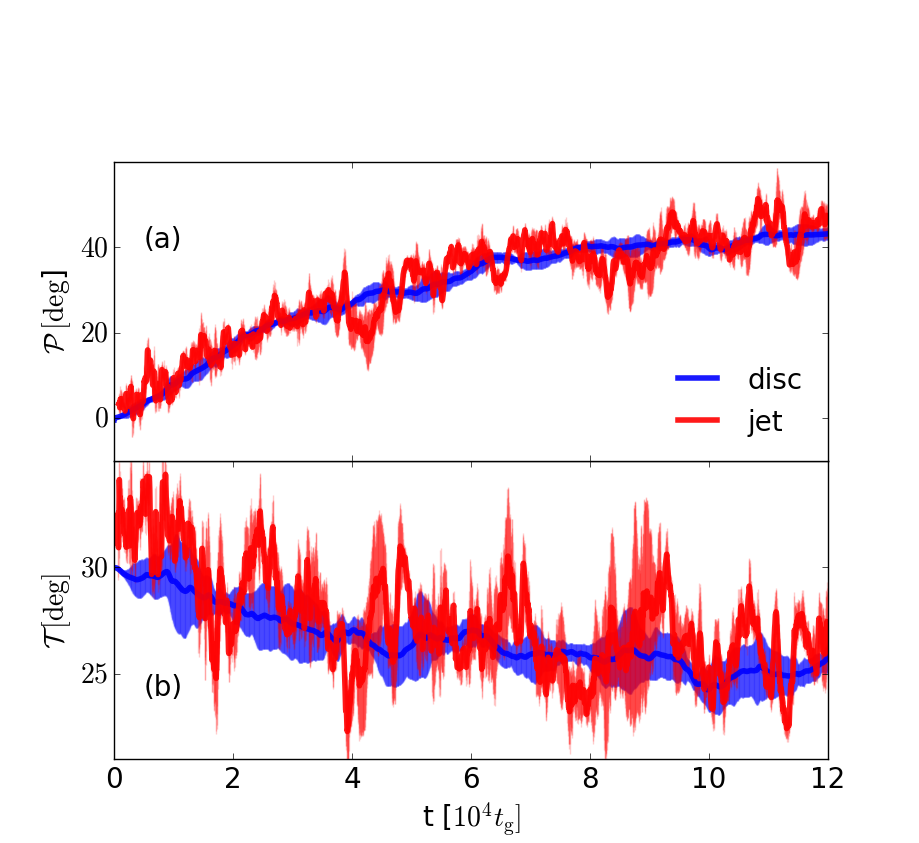} 
\caption{Time-dependence of precession [panel (a)] and tilt [panel (b)]
  angles of the disc (blue) and the
  upper jet (red), as measured at
  $50 \: r_g\le r \le 150 \: r_g$ in model W-R, with $1\sigma$ error bars showing variation across this radius interval. The jet direction
  closely tracks the disc
  rotation axis. Jet wobbling by several degrees, evident in
  both panels, could give rise to jet-powered high-energy flares. }
\label{fig:tiltprec}
\end{figure}

Figures~\ref{fig:visual} and~\ref{fig:tiltprec}(a) show that the jets precess together with the disc: the 
spatially-averaged (over $50r_g\le r\le 150r_g$) $\mathcal{P}$ of the disc
and jet closely track each other in time. %
Figure~\ref{fig:tiltprec}(b) shows that the disc and jet not only
precess together but also align together, with the tilt angle
decreasing from $\sim 30^\circ$ to $\sim 25^\circ$. %

\begin{figure}
\includegraphics[width=1.0\linewidth,trim=0 0 0 85,clip=true]{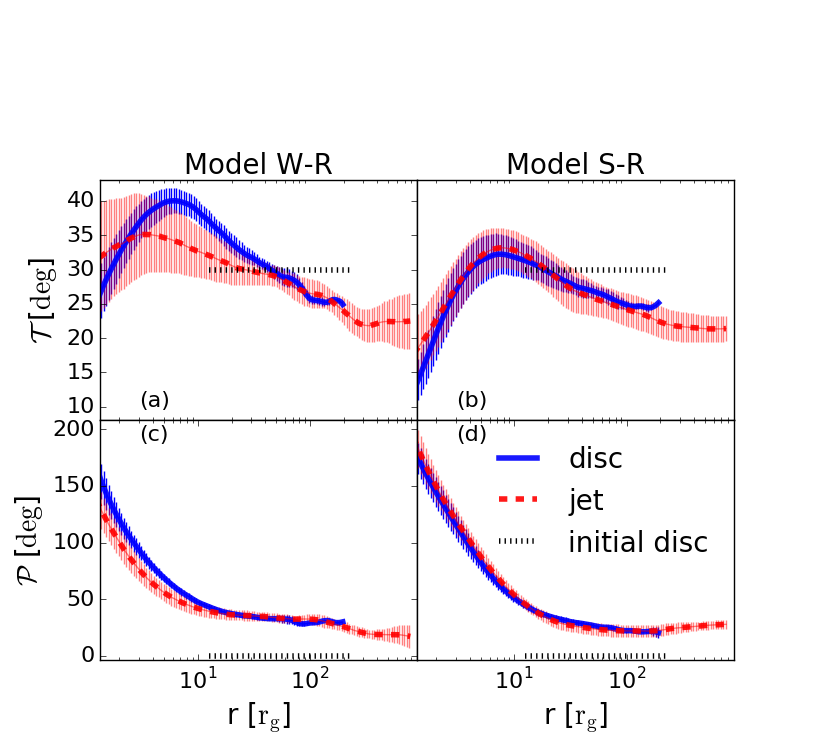} 
\caption{
  Radial profiles of tilt $\mathcal{T}$ and precession
  $\mathcal{P}$ angle of an upper jet (red dotted
  lines; the other jet is similar), averaged over
  $t=4.5-5.5\times10^4t_g$, closely follow those of the disc (blue curves) at
  $r\gtrsim 3r_g$ (using spherical radius for both the jet and the
  disc, error bars indicate $1\sigma$-variation across this time
  interval). In both cases of weak (model W-R, left panels) and
  strong (model S-R, right panels) magnetic flux, jets
  propagate along the disc rotational axis, not the BH spin axis. At
  $r\lesssim10r_g$, radial tilt oscillations cause variations in the
  disc and jet tilt angles [panels (b), (d)]. If
  strong magnetic flux is present, as in model S-R [panel (d)], it can
  bend the innermost $\lesssim3r_g$ of the disc and jets into partial
  alignment with the BH
  spin. %
}
\label{fig:tiltprecr}
\end{figure}

Figure~\ref{fig:tiltprecr} shows the tilt $\mathcal T$ (top row) and precession $\mathcal P$ (bottom row) 
angles for models W-R (left column)
and S-R (right column) vs $r$ at $t = 5 \times 10^4\: t_g$. 
The jets align with the disc far
from the BH. Radial tilt oscillations lead to
the peak in disc tilt angle at $r\sim 10~r_g$, as expected for a
thick disc \citep{lubow00,lubow02}, in agreement with previous work
\citep{fragile07}. Strong twisting close to the BH, indicated by the
steep radial dependence of disc precession angle at $r\lesssim
10~r_g$, builds up early on in the simulation and remains
\hbox{constant thereafter
\citep{fragile07}.}

Figure~\ref{fig:tiltprecr}(a-b) shows that in model S-R the disc and
jets tend to align with the black hole spin in the inner
${\rm few}\, r_g$, while in the weak field case W-R the jet remains
misaligned up to the event horizon. This result supports the notion that strong jets
can align the inner parts of disc-jet systems with the BH spin
\citep{mtb13,polko17}. However, 
the alignment can be very limited
in distance, $r\lesssim 3r_g$, and
magnitude: the stronger
magnetic flux in model \hbox{S-R} leads to an additional disc
alignment relative to model \hbox{W-R} by only $\sim4$ degrees at
$r\sim15{-}50r_g$. Also, the LT torque acting on a twisted accretion
disc \highlight{can contribute to the alignment indirectly. When
  precessed material mixes due to viscous dissipation \citep[see,
  e.g., ][]{king05, sorathia13}, the net result is alignment with the
  BH spin. Since jets and (MRI-driven) dissipation both depend on
  magnetic field strength, both could lead to differences in alignment between S-R and W-R. } 

\section{Discussion and Conclusions} \label{sec:Discussion} 

We carried out GRMHD simulations of tilted BH accretion discs at
the highest resolutions to date (effective resolution $\sim$1 billion cells). We find
that our discs, with an initial tilt of $30$ degrees relative to the
BH of spin $a=0.9375$, undergo LT solid-body--like precession (see
also \citealt{fragile07}). Our simulations for the first time show
that tilted precessing discs can launch relativistic jets. The jets
(i)~propagate along the disc rotation axis and (ii)~precess together
with the disc. 

\highlight{The amount of large-scale vertical magnetic flux significantly affects the
  orientation of the disc and jets: when the BH is saturated with
  the flux, as in our model \hbox{S-R}, it is able to warp the
  disc and jets into partial alignment with the BH spin in the inner
  few $r_g$ [Figure~\ref{fig:tiltprecr}(b)].  Differences in magnetic
  flux content can also
  explain differences with \citet{mtb13}, who reported partial
  alignment of the jet with the BH spin up to $\sim$$100$ $r_{g}$. We have replicated  in H-AMR one
  of their models (A0.99N100T0.6). Their initial conditions
  contain much more magnetic flux in a 100-fold
  larger disc.
  As accretion
  drags the large-scale vertical magnetic flux inwards over time, not only the BH but
  also the inner accretion disc gets saturated with magnetic
  flux (i.e., reaches the MAD state), with average $p_g/p_{mag}$
  within $100$ $r_g$ decreasing to $\sim$$1$, much lower than $18$
  and $35$ in our models S-R and W-R, respectively. 
}

Type-C QPOs observed from BH XRBs have
been interpreted as LT precession of the inner accretion flow
\citep{stella98, ingram09, ingram16}. 
If the jet base is X-ray bright \citep{markoff05}, the
precession can lead to quasi-periodic swings in the X-ray polarisation
angle in addition to those expected from the inner disc alone
\citep{ingram15}. This can be tested with the \textit{Imaging X-ray
  Polarimetry Explorer} (\textit{IXPE}) due to launch in
2020.  Observing longer-term changes in large-scale
jet orientation due to LT precession
\citep[e.g.,][]{ekers78,bridle79,vidal11,kalamkar16}
can enable new tests of GR, BH accretion, and jet physics
\citep[e.g.,][]{2012PhRvL.108f1302S}.  The precession of AGN jets can cause
them to spread their power over a large area and heat the ambient gas
instead of escaping out of the galaxy/cluster
(\citealt{nawaz16,yang16}).
Together with the jet wobbling
due to magnetic instabilities \citep{tchekhovskoy16}, this can provide
an explanation for the unexpectedly high temperature of the
intra-cluster medium, known as the cooling flow problem.

In addition to solid-body--like precession, we find that tilted
disc-jet systems align with the BH spin axis over longer, accretion
timescales \highlight{(see also \citealt{foucart11})}. Strong large-scale magnetic fields and the associated
powerful jets accelerate this alignment
[Fig.~\ref{fig:tiltprecr}(a-b)]. 
Short-timescale wobbles of several degrees in jet orientation
superimposed on the smooth precession and alignment trends
[Figure~\ref{fig:tiltprec}(a-b)] could boost jet emission in and out of
our line-of-sight and result in jet-powered high-energy flares similar
to those in the X-ray light curve of a tidal disruption event Swift
J1644 \citep[e.g.,][]{Bloom2011}.

We found it important for the simulations to resolve the MRI
throughout the disc \highlight{(see also \href{https://www.youtube.com/watch?v=fFQyYag_8QA&index=7&list=PL39mDr1uU6a5RYZdXLAjKE1C_GAJkQJNv}{this 3D visualization} of the effects of resolution on our weak-field models)}. Magnetic turbulence causes angular momentum transport and
radial disc expansion [Fig.~\ref{fig:time}(i-j)] that substantially
slows down the precession [Fig.~\ref{fig:time}(e-f)] and alignment
[Fig.~\ref{fig:time}(g-h)] by redistributing most of the angular
momentum out of reach of the BH. Our thick discs
are isolated, and external disc feeding from large radii (e.g., by a
stellar debris stream in TDEs or a geometrically thin disc in AGN
and XRBs) can affect the disc expansion either by the addition of new gas
or by applying external pressure. Also, geometrically-thin discs ($H/R<\alpha$) have larger viscosity, which can reduce disc expansion and affect precession and alignment. We use H-AMR to study that limit in future work.

\section{Acknowledgments}
We thank Chris Fragile, Nick Stone, and Asaf Pe'er for helpful comments and Mark Vanmoer for his help with visualization. This research was made possible by NSF PRAC award no.~1615281 at the
Blue Waters sustained-petascale computing project and supported in part under grant no.~NSF~PHY-1125915.
ML and MK were supported
by the Netherlands Organisation for Scientific Research (NWO)
Spinoza Prize, CH by the Amsterdam Science Talent Scholarship, AI by the NWO VENI grant
(no.~639.041.437), AT by the TAC and NASA Einstein (grant no.~PF3-140131) postdoctoral fellowships, and SM by the NWO VICI grant (no.~639.043.513).

\section{Supporting Information}
Additional Supporting Information may be found in the online version
of this article: code technical details and tests, and movie files. See our \href{https://www.youtube.com/playlist?list=PL39mDr1uU6a5RYZdXLAjKE1C_GAJkQJNv}{YouTube
  playlist} for 3D visualizations of all five models.

\label{sec:acks}

{\small
\bibliography{mybib,sasha,newbib}
\bibliographystyle{mn2e}
}
\label{lastpage}
\end{document}